\documentclass[preprint,aps,superscriptaddress,showpacs]{revtex4}
\usepackage{amssymb}
\usepackage{amsmath}
\usepackage{graphicx}
\usepackage[normalem]{ulem}
\usepackage[dvips]{color}
\usepackage{array,longtable,multirow}
\begin{document}

\title{Half-lives of  Double $\beta ^+$-decay with  Two Neutrinos }

\author{Yuejiao Ren}
\affiliation{Department of Physics, Nanjing University, Nanjing
210093, China}
\author{Zhongzhou Ren\footnote{Corresponding author, zren$@$nju.edu.cn}}
\affiliation{Department of Physics, Nanjing University, Nanjing
210093, China}
\affiliation{Center of Theoretical Nuclear Physics,
National
 Laboratory of Heavy-Ion Accelerator, Lanzhou 730000, China}

\begin{abstract}
Nuclear double $\beta ^-$-decays with two neutrinos were observed
for many years and a systematic law describing the relation between
their half-lives and decay energies was also proposed recently
[Phys. Rev. C89, 064603 (2014)]. However, double $\beta ^+$-decay
($\beta ^+\beta^+)$ with emission of both two positrons and  two
neutrinos has not been observed up to date. In this article, we
perform a systematic analysis on the candidates of double $\beta
^+$-decay, based on the 2012 nuclear mass table. Eight nuclei are
found to be the good candidates for double $\beta ^+$-decay and
their half-lives are predicted according to the generalization of
the systematic law to double $\beta ^+$-decay. As far as we know,
there is no theoretical result on double $\beta ^+$-decay of nucleus
$^{154}Dy$ and our result is the first prediction on this nucleus.
This is also the first complete research on eight double $\beta
^+$-decay candidates based on the available data of nuclear masses.
It is expected that the calculated half-lives of double $\beta
^+$-decay in this article will be useful for future experimental
search of double $\beta ^+$-decay.

\

Key words : double $\beta ^+$-decay with two neutrinos,  half-lives,
systematic law, weak interaction.

\end{abstract}

\pacs{23.40.-s, 23.40.Bw }

\maketitle

\section{introduction}
Nuclear double  $\beta $-decay with two neutrinos is a rare and
exotic process occurring in  nuclei with long lifetimes. Since
Goeppert-Mayer predicted that there is  double   $\beta $-decays
 in 1935 \cite{may}, many researches have been carried out
 on double
 $\beta$-decay with two neutrinos or without neutrinos
\cite{hax,eji,ack,gan,arn,bar,saa,aud1,aud2}. Plenty of results  can
be found in the references of  the articles
\cite{moe,kla,rad,avi,ago,fae1,fae2,cau,suh1,suh2,suh3,vog,ver}.
Although much experimental effort has been devoted to double
 $\beta$-decay, only the half-lives of eleven nuclei with double $\beta ^-$-decays
 were obtained and the half-life of a single nucleus with double
 electron capture (ECEC) was measured due to the difficulty of experimental detection
 in this field \cite{saa}.  On double $\beta ^+$ decay ($\beta ^+
 \beta ^+$), experimental data of definite half-lives are still not available \cite{saa}.
Theoretically there are many successful calculations on
   the double $\beta ^-$-decay half-lives
   \cite{hax,eji,ack,kla,rad,avi,ago,fae1,fae2,cau,suh1,suh2,suh3,vog,ver}.
   However, calculations on  the double $\beta ^+$-decay half-lives are much less
   as they are compared with those on double $\beta ^-$-decay
   \cite{hax,eji,ack,kla,rad,avi,ago,fae1,fae2,cau,suh1,suh2,suh3,vog,ver}.
   Haxton et  al. estimated the half-lives of six nuclei from
   mass number A=78 ($^{78}Kr$) to mass number A=136 ($^{136}Ce$)
   and showed that their half-lives for double $\beta ^+$-decay
   ranged approximately from  $10^{26}$ years
   to $10^{31}$ years \cite{hax}. Since the calculations from Haxton et al. \cite{hax},
    thirty years have passed and the data of nuclear masses have been updated significantly
\cite{hax,aud1,aud2}.
   Therefore, it is necessary to calculate the half-lives of double $\beta
   ^+$-decay with new nuclear mass tables because the half-lives
   are very sensitive to the decay energies.
    Recently, Suhonen made the theoretical
   investigation of  double $\beta ^+$-decay for two nuclei,
   $^{78}Kr$ \cite{suh1} and  $^{96}Ru$ \cite{suh2}.
As definite double $\beta ^+$-decay half-lives are not available in
experiments now, it is interesting to make a complete calculation on
the possible candidates of double $\beta ^+$-decay and to predict
new candidates for future experiments.

\section{Numerical results and discussions on the  double $\beta ^+$-decay
half-lives}

  Based on the 2012 nuclear mass table \cite{aud1,aud2}, we calculate
the decay energy of  all possible double $\beta ^+$-decay emitters
and pay special attention to the parent nuclei where they are
denoted with a symbol ($2\beta ^+\,?$) in the table. The number of
possible double $\beta ^+$-decay emitters is 40 in the mass table
\cite{aud1,aud2} and their mass number ranges from A=36 ($^{36}Ar$)
to A=252 ($^{252}Fm$) \cite{aud1,aud2}. In accordance with textbooks
and published articles,  the  decay energy of double $\beta
^+$-decay is defined as
\begin{eqnarray}\label{eq1}
Q_{2\beta } (MeV)\,=\,[\,M(A,Z)\,-\,M(A,Z-2)\,-\,4\,M_e\,]\,C^2
\end{eqnarray}
 where M(A,Z) is the mass excess of the parent nucleus and M(A, Z-2)
 is the mass excess of the daughter nucleus. $M_e$ is the rest mass of
 an electron (or a positron) and $C$ is the speed of light in vacuum
 ($C=1$ in natural units).
 Our numerical calculations for the 40 possible emitters shows that
 many of them have negative decay energy to double $\beta ^+$-decay
 or have approximately zero decay energy. Only eight of them have
 significantly positive energies for double $\beta ^+$-decay.
 We list the eight emitters (parent nuclei) and their daughter nuclei
 in Table 1, together with the decay energy and the isotopic abundance (IS) of
 parent nuclei (or their $\alpha$-decay
half-lives ($T^{\alpha }$) when $\alpha $-decay  is observed).

\begin{table}[htb]
\small \centering \caption {The decay energy of double $\beta
^+$-decay candidates from the 2012 nuclear mass table
 where the decay energy is defined as $Q_{2\beta
}\,=\,[\,M(A,Z)\,-\,M(A,Z-2)\,-\,4\,M_e\,]\,C^2$ (MeV).  We also
list the isotopic abundance (IS) of parent nuclei or their
$\alpha$-decay half-lives ($T^{\alpha }$) when $\alpha $-decay  is
observed. The experimental data of  nuclear masses and isotope
abundance are from references \cite{aud1,aud2}.} \label{tab1}
\renewcommand{\tabcolsep}{2mm}
\begin{tabular}{cccccc}
\hline \hline Parent
        & Daughter&M(A,Z) (MeV)&M(A,Z-2) (MeV)
         &$Q_{2\beta}$(MeV)
        &IS or T$^{\alpha }$\\
\hline
$^{78}$Kr & $^{78}$Se &  -74.180&-77.026&0.802&IS=0.355\% \\
$^{96}$Ru & $^{96}$Mo &  -86.079&-88.794&0.671&IS=5.54\% \\
$^{106}$Cd & $^{106}$Pd &  -87.132&-89.907&0.731&IS=1.25\% \\
$^{124}$Xe & $^{124}$Te &  -87.661&-90.525&0.820&IS=0.095\% \\
$^{130}$Ba & $^{130}$Xe &  -87.262&-89.880&0.574&IS=0.106\% \\
$^{136}$Ce & $^{136}$Ba &  -86.509&-88.887&0.334&IS=0.185\% \\
$^{148}$Gd & $^{148}$Sm &  -76.269&-79.336&1.023&T$^{\alpha }$=70.9 y \\
$^{154}$Dy & $^{154}$Gd &  -70.394&-73.705&1.267&T$^{\alpha }$=3.0 My \\
\hline \hline
\end{tabular}
\end{table}

   Table 1 shows that all eight parent nuclei have
positive decay energies for double $\beta ^+$-decay. From $^{78}$Kr
to $^{136}$Ce, they are naturally occurring isotopes and their decay
energy  is lower than 1 MeV. For nuclei $^{148}$Gd and $^{154}$Dy,
they are unstable for $\alpha $-decay but their $\alpha $-decay
half-lives are long (T$^{\alpha }$=70.9 years (y) or T$^{\alpha
}$=3.0 My) \cite{aud1,aud2}.  Especially their decay energies for
double $\beta ^+$-decay are higher than 1 MeV. So the last two
nuclei ($^{148}$Gd and $^{154}$Dy) could be very interesting to
observe double $\beta ^+$-decay because decay half-lives are very
sensitive to decay energies.

   For the purpose of calculating the double $\beta ^+$-decay half-lives of
   the above eight nuclei, we will generalize the  systematic law of
 double  $\beta ^-$-decay half-lives \cite{ren8} to the
 case of double $\beta ^+$-decay. It is proposed  \cite{ren8} that
 there is a new systematic law between the half-lives and decay
 energies for  double  $\beta ^-$-decay with two neutrinos
\begin{eqnarray}\label{eq2}
lg\,T_{1/2} (Ey)\,=\,(a\,-2\,lg(2\,\pi
\,Z/137)\,+\,S\,)\,/\,{Q_{2\beta} (MeV)}
\end{eqnarray}
where the constant $a$ is obtained by fitting the experimental data
of double  $\beta ^-$-decay and its value is $a=5.843$ \cite{ren8}.
The physical meaning of $a$ is related to the square of the strength
of the weak interaction which leads to the instability of a nucleus.
$Z$ is the charge number of the parent nucleus and the second term
in the right side of Eq. (2) is the effect  of the Coulomb field on
double  $\beta ^-$-decay half-lives due to the emission of two
electrons \cite{ren8}. $S=2$  when the neutron number of parent
nuclei is a magic number and \,$S=0$ when the neutron number of
parent nuclei is not a magic number \cite{ren8}.

    In order to generalize the law of Eq.(2) from double  $\beta
    ^-$-decay to that of double $\beta ^+$-decay, we analyze the
    contribution of the Coulomb field on half-lives.
    It is well known from textbooks \cite{pre,sha,wa1} and from
    previous researches \cite{ni6,ni7,li6,li14,zha,mol}
    that the effect of the Coulomb field
    in  single $\beta ^+$-decay is oppositive as compared with that
     in  single $\beta ^-$-decay.  The effect of the Coulomb field
     will shorten the half-lives of a $\beta ^-$-decay but it will
     prolong the half-lives of a $\beta ^+$-decay when the decay
     energy is fixed.  This is due to the oppositive charge between
     a positron and an electron. These are also
     confirmed by numerical calculations of $\beta $-decay
     half-lives \cite{ni6,ni7,li6,li14,zha,mol}.  It is natural to
      extend this idea from
     single  $\beta $-decay to double $\beta $-decay. Therefore,
     the systematic law of double $\beta ^+$-decay half-lives is
\begin{eqnarray}\label{eq3}
lg\,T_{1/2} (Ey)\,=\,(a\,+2\,lg(2\,\pi \,Z/137)\,)\,/\,{Q_{2\beta}
(MeV)}
\end{eqnarray}
Here the parameter $a$ of eq.(3) is the same value as that in
Eq.(2). The second term of eq.(3) is the same value  as that in
Eq.(2) but with oppositive contribution due to different signs of
charges between positron and electron.  In Eq. (2) the last term $S$
 is from the contribution of neutron closed shell when the neutron number of
 parent nucleus is a magic number for double $\beta ^-$-decay.
 However, for double $\beta ^+$-decay, it is difficult to occur due
 to the requirement of decay energy and there are only eight suitable candidate
 nuclei  in Table 1.  The proton number of the
 eight nuclei in Table 1 is not a magic number and therefore it is natural
 to choose $S=0$.  Finally we obtain the analytical formula Eq.(3) to
 calculate the half-lives of double $\beta ^+$-decay, without
 introducing new adjusting parameters. So it can be concluded that
 the number of adjusting parameters in this article is  the minimum
 , with the same spirit as that in our previous research
 \cite{ren8}.

    We use Eq.(3) to calculate the half-lives of double $\beta ^+$-decay
    and the numerical results are  drawn in Figure 1 and
  listed in Table 2. In
  Fig.1, the X-axis is the decay energy (MeV) and the Y-axis is the
  logarithm of the double $\beta ^+$-decay half-lives (Ey) where
  $Ey=10^{18}$ years.   Fig.1 shows that the calculated
  half-lives are very sensitive to the variation of the decay energies.
    Although decay energies vary in a narrow range from 0.334 MeV
    ($^{136}$Ce ) to 1.267 MeV ($^{154}$Dy ), the change to the half-lives
    is approximately a factor of $10^{15}$. This  strongly suggests
   that double $\beta ^+$-decay with a higher decay energy is better
   for observation. We recommend nuclei $^{154}$Dy
   and $^{148}$Gd as good candidates to observe double $\beta
   ^+$-decay.

    Now we  make a detailed
  discussion on the calculated results of Table 2.
  In Table 2,  Column 1 represents the  parent nuclei and column 2 represents
  experimental decay energy.  Column 3 denotes   the logarithm of the calculated
  double $\beta^+$-decay half-lives with equation (3). Column 4 is
  our calculated
  double $\beta^+$-decay half-lives, which is convenient for
  comparison with future experiments.
The fifth column provides the calculated results from Suhonen
\cite{suh1,suh2} and  they are also listed for comparison.
 It is  seen
from the comparison of our results (column 4) and other results (the
last column) that the two sets of theoretical half-lives  are close
for nuclei $^{78}$Kr and $^{96}$Ru although different calculation
approaches were used. There was an old calculation \cite{kla} in
1991 on seven nuclei but it is not suitable for comparison with our
results because  the experimental data of nuclear masses have been
updated twice. This is related  to different inputs of decay
energies between the old calculation and that of this manuscript. So
 reasonable agreement is
reached for different calculations. This is good for future
experiments. It is widely accepted that different approaches are
useful for the further development of the field. From the last two
rows of  Table 2, it is seen that the half-lives of $^{148}Gd$ and
$^{154}$Dy are shorter due to their higher double $\beta^+$-decay
energies. Because both $^{148}$Gd and $^{154}$Dy have been observed
to have $\alpha $-decay
 and their $\alpha $-decay half-lives are long
(T$^{\alpha }$=70.9 years  for $^{148}Gd$ and  T$^{\alpha }$=3.0 My
for $^{154}$Dy ) \cite{aud1,aud2,ren6,qian} , it is interesting to
detect the double $\beta^+$-decay from the two nuclei.  We can
estimate a branch ratio \cite{aud1,aud2,ren6} from the half-lives of
double $\beta^+$-decay (Table 2) and $\alpha $-decay (Table 1). The
branching ratio (BR) between double $\beta^+$-decay and $\alpha
$-decay is $BR\,=\,T^{\alpha }/T_{1/2}(theor)\,=\,(3\times
10^6)\,/\,(2.35\times 10^5\times 10^{18})=1.28\times 10^{-17}$ for
$^{154}$Dy. This value can be used for reference of future
experiments to search the double $\beta^+$-decay of $^{154}$Dy.

\begin{table}[htb]
\small \centering \caption {The  double $\beta ^+$-decay half-lives
of even-even  isotopes calculated with new systematic law
($T_{theor}$) and the corresponding logarithms ($lgT_{theor}$). The
units of the half-lives are Ey ($10^{18}$ years). The experimental
decay energies of nuclei ($Q_{2\beta}$ (MeV)) are also listed in the
table. The calculated half-lives from another group \cite{suh1,suh2}
are listed in the last  column for comparison.} \label{tab2}
\renewcommand{\tabcolsep}{2mm}
\begin{tabular}{ccccc}
\hline \hline Nuclei
        &$Q_{2\beta}$(MeV)
        &$lgT_{1/2}(theor)$
 &$T_{1/2}(theor)$ (Ey) &$T_{1/2}(other)$ (Ey) \\
\hline $^{78}$Kr & 0.802& 7.828& $6.73\times
10^7$&$(4.94-15.8)\times 10^{7}$\cite{suh1}\\
$^{96}$Ru & 0.671& 9.616& $4.13\times 10^9$ & $(1.2-10)\times
10^8$\cite{suh2}\\
$^{106}$Cd & 0.731& 8.930& $8.51\times 10^8$      &\\
$^{124}$Xe & 0.820& 8.086& $1.22\times 10^9$    &
\\
$^{130}$Ba & 0.574& 11.606& $4.04\times 10^{11}$    &
\\
$^{136}$Ce & 0.334& 20.037 & $1.09\times 10^{20}$  &
\\
$^{148}$Gd & 1.023 & 6.625& $4.22\times 10^6$ &
\\
$^{154}$Dy & 1.267  &5.371 & $2.35\times 10^5$    &  \\
\hline \hline
\end{tabular}
\end{table}

\begin{figure}[htb]
\centering
\includegraphics[width=12cm]{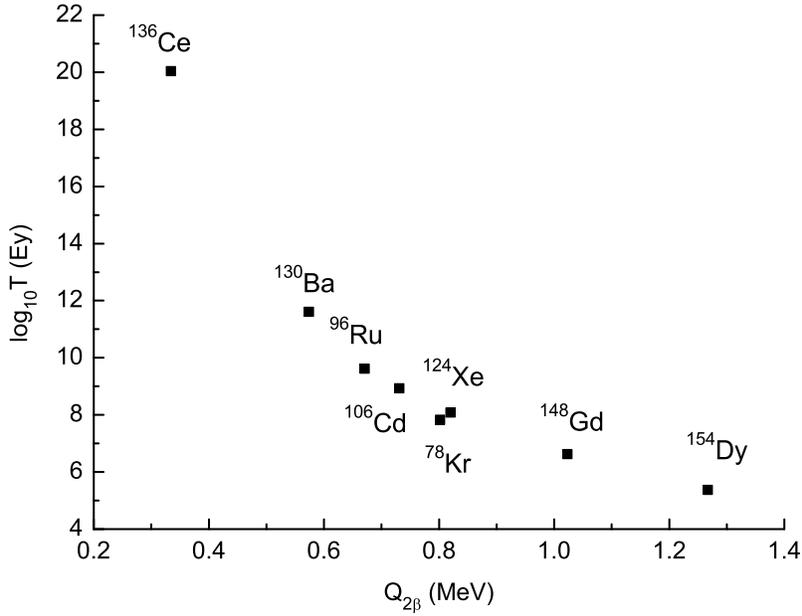}
\caption{ (Color online) Logarithms of the theoretical double $\beta
^+$-decay half-lives  for ground-state transitions of eight
even-even parent nuclei from $^{78}$Kr to $^{154}$Dy. The units of
the half-lives are Ey ($10^{18}$ years ). X-axis is the decay energy
(MeV). The half-lives of double $\beta ^+$-decay of nuclei
$^{148}$Gd and $^{154}$Dy are shorter than other nuclei and they
could be very interesting for future experimental observation of
double $\beta ^+$-decay. }
\end{figure}

\section{Summary}

In summary, we collect the experimental data of possible double
$\beta ^+$-decay nuclei based on the 2012 nuclear mass table and
find that eight nuclei have significantly positive decay energies. A
systematic calculation of double $\beta ^+$-decay half-lives for the
eight nuclei is carried out with the  analytical formula being a
natural generalization of the systematic law from  double $\beta
^-$-decay to double $\beta ^+$-decay. Numerical results show that
 the half-lives of
double $\beta ^+$-decay of nuclei $^{148}Gd$ and $^{154}Dy$ are
shorter than other nuclei and they can be very interesting for
future experimental observation of double $\beta ^+$-decay. The
branching ratio (BR) between   double $\beta^+$-decay and $\alpha
$-decay is also estimated  for $^{154}$Dy. Our result on nucleus
$^{154}Dy$ is the first prediction as far as we know. This is a
complete calculation on eight double $\beta ^+$-decay nuclei, which
will be useful for future experimental researches \cite{cao1}.

\section{Acknowledgements}

This work is supported by National Natural Science Foundation of
China (Grant Nos. 11035001, 10975072, 11120101005, 11175085,
11375086), by the Priority Academic Program Development of Jiangsu
Higher Education Institutions.

\end{document}